\documentclass[twocolumn,trackchanges]{aastex631}

\begin{document}

\title{A Highly Magnetic Ultra Massive White Dwarf with a 23-minute Rotation Period}

\correspondingauthor{Jincheng Guo; Xiaofeng Wang}
\email{Andrewbooksatnaoc@gmail.com; wang\_xf@mail.tsinghua.edu.cn}

\author[0000-0002-8321-1676]{Jincheng Guo}
\affiliation{Department of Scientific Research, Beijing Planetarium, Xizhimenwai Road, Beijing 100044, China }

\author{Xiaofeng Wang}
\affiliation{Physics Department, Tsinghua University, Beijing 100084, China }

\author{Qichun Liu}
\affiliation{Physics Department, Tsinghua University, Beijing 100084, China }

\author[0000-0003-3460-0103]{Alexei V. Filippenko}
\affiliation{Department of Astronomy, University of California, Berkeley, CA 94720-3411, USA }

\author[0000-0001-5955-2502]{Thomas G. Brink}
\affiliation{Department of Astronomy, University of California, Berkeley, CA 94720-3411, USA }

\author{Jingkun Zhao}
\affiliation{Key Laboratory of Optical Astronomy, National Astronomical Observatories, Chinese Academy of Sciences, Beijing 100101, China }

\affiliation{School of Astronomy and Space Science, University of Chinese Academy of Sciences, Beijing 100049, China}

\author[0000-0002-2636-6508]{WeiKang Zhang}
\affiliation{Department of Astronomy, University of California, Berkeley, CA 94720-3411, USA }

\author[0000-0002-6535-8500]{Yi Yang}
\affiliation{Physics Department, Tsinghua University, Beijing 100084, China }

\author{Jie Lin}
\affiliation{CAS Key laboratory for Research in Galaxies and Cosmology, Department of Astronomy, University of Science and Technology of China, Hefei, 230026, China}
\affiliation{School of Astronomy and Space Sciences, University of Science and Technology of China, Hefei 230026, China}

\author{Haowei Peng}
\affiliation{Physics Department, Tsinghua University, Beijing 100084, China}

\author{Hailiang Chen}
\affiliation{Yunnan observatories, Chinese Academy of Sciences (CAS), Kunming, 650216, China}

\author{Davron O. Mirzaqulov}
\affiliation{Ulugh Beg Astronomical Institute, Astronomy Street 33, Tashkent 100052, Uzbekistan }

\author{Shuhrat A. Ehgamberdiev}
\affiliation{Ulugh Beg Astronomical Institute, Astronomy Street 33, Tashkent 100052, Uzbekistan }
\affiliation{Samarkand State University, 15, University Boulevard, 140104, Samarkand, Uzbekistan }

\author{Bin Ma}
\affiliation{School of Physics and Astronomy, Sun Yat-sen University, Zhuhai 519082, China }
\affiliation{CSST Science Center for the Guangdong-Hong Kong-Macau Greater Bay Area, Zhuhai 519082, China }

\author{Jun Mo}
\affiliation{Physics Department, Tsinghua University, Beijing 100084, China }

\author{Cheng Liu}
\affiliation{Department of Scientific research, Beijing Planetarium, Xizhimenwai Road, Beijing 100044, China }

\author{Gaobo Xi}
\affiliation{Physics Department, Tsinghua University, Beijing 100084, China }

\author{Xiaojun Jiang}
\affiliation{Key Laboratory of Optical Astronomy, National Astronomical Observatories, Chinese Academy of Sciences, Beijing 100101, China  }
\affiliation{School of Astronomy and Space Science, University of Chinese Academy of Sciences, Beijing 100049, China}

\author{Danfeng Xiang}
\affiliation{Department of Scientific research, Beijing Planetarium, Xizhimenwai Road, Beijing 100044, China }

\author{Jicheng Zhang}
\affiliation{School of Physics and Astronomy, Beijing Normal University, Beijing 100875, China}

\begin{abstract}

We present a physical characterization of TMTS J00063798+3104160 (J0006), a rapidly rotating, ultra-massive white dwarf (WD) identified in high-cadence light curves from the Tsinghua University–Ma Huateng Telescope for Survey (TMTS). A coherent 23-minute periodicity is detected in TMTS, {\it TESS}, and ZTF photometry. A time series of low-resolution spectra with the Keck-I 10\,m telescope reveals broad, shallow hydrogen absorption features indicative of an extreme magnetic field and shows no evidence for radial-velocity variations. Atmospheric modeling yields a magnetic field strength of $\sim$\,250\,MG, while {\it Gaia} astrometry and photometry imply a mass of $1.06 \pm 0.01$\,M$_{\odot}$. A significant infrared excess is detected in the {\it WISE} $W1$ band and is well fitted by a 550\,K blackbody, likely arising from residual material of a merger. We interpret the 23-minute photometric modulation as the rotation period of an isolated, massive WD formed likely through the merger of a double WD binary. With one of the shortest rotation periods known among candidate merger remnants and with constraints from a deep {\it Einstein Probe} X-ray nondetection, J0006 provides a rare and important observational window into the poorly explored intermediate stages of post-merger evolution.

\end{abstract}

\keywords{Stellar astronomy: individual (TMTS J00063798+3104160): Compact objects: White dwarf stars: DA stars Variable stars: Magnetic variable stars.}

\section{Introduction}
Single white dwarfs (WDs) are generally considered nonvariable sources owing to their lack of internal energy generation and relatively simple atmospheric structure. However, as early as 1968, HL Tau 76 was identified as the first intrinsically variable (pulsating) WD \citep{Landolt1968}. Since then, an increasing number of variable single WDs have been discovered. With the advent of wide-field, high-candence, and high-precision time-domain surveys, the known population and phenomenological diversity of variable WDs have expanded rapidly and systematically \cite[{\it Kepler, K2, TESS}, ZTF, LSST, \it{PLATO}] {Borucki2010,Howell2014,Ricker2014,Bellm2019,Ivezic2019,Rauer2014}. To date, the observed variability in WDs has been primarily attributed to stellar pulsations, transits, rotation, and gravitational lensing \citep{Winget2008,Fontaine2008,Corsico2019,Vanderburg2015,Jewett2024,Kruse2014}. The rotation of isolated WDs can be probed through asteroseismology, spot modulation, rotational broadening of the Ca~II~K line, and high-resolution spectroscopy of the H$\alpha$ core in hydrogen-rich WDs \citep{Bognar2024,Rosa2024,Berger2005}. In isolated WDs, surface spots, thought to arise from magnetic-pressure-induced distortions \citep{Fendt2000}, are commonly invoked to explain photometric variability with periods ranging from several hours to about a day, consistent with typical WD rotation timescales \citep{Rosa2024}. 

WDs with rotation periods shorter than 30\,min are quite rare, with only a handful of such objects having been identified thus far. Among them, RE J0317-853 was discovered by \cite{Barstow1995}, exhibiting a rotation period of 725.4\,s and hosting a strong magnetic field of $\sim 340$\,MG. \cite{Reding2020} presented an isolated WD with a 317\,s rotation period and prominent magnetic emission. \cite{Caiazzo2021} reported the detection of a highly magnetized and rapidly rotating WD with a period of only 6.94\,min. In particular, an isolated WD with only a 70\,s spin period (the shortest known) was reported \citep{Kilic2021}. EGGR 156 was reported by \cite{Williams2022} with a 22.4\,min rotaion period and a mass of 1.298\,$M_{\odot}$. Most recently, an interesting WD with a rotation period of 6.6\,min, magnetic field strength $\sim 400$--600\,MG, and mass  1.12\,M$_{\odot}$ was detected in an open cluster, and it shows variable double-peaked emission lines \citep{Yan2025,Cristea2025}. \cite{Williams2025} reported two more WDs with rotation periods of 84 and 169\,s. These rapidly spinning WDs are typically detected through photometric modulations caused by surface spots, which are thought to be generated by the WD's strong magnetic field. However, the origin of strong magnetic fields in WDs remains an open question. The proposed mechanisms include fossil fields inherited from the progenitor stars, dynamo processes during the late stages of stellar evolution, and magnetic field amplification through binary interactions or mergers \citep{Camisassa2024,Ferrario2020,Castro2025}. 

Observational studies have shown that magnetic WDs exhibit a wide range of magnetic field strength. The field strength and geometry measurements are obtained primarily through model fitting of the Zeeman splitting in WD spectral lines \citep{Kepler2013}. On the other hand, investigations of the polarization properties of magnetic WDs provide additional insights into their magnetic field structure and the physical processes operating in their atmospheres \citep{Landstreet2015}. 

By searching for short-period variable WDs among sources common to the Tsinghua University–Ma Huateng Telescope for Survey  (TMTS) \citep{Zhang2020} and the {\it Gaia} DR3 WD catalog \citep{Gentile2021}, several special variable WDs have been discovered. Among these, two pulsating DA WDs were studied in detail \citep{Guo2023,Guo2024}. Additionally, a unique WD, TMTS J00063798+3104160 (hereafter J0006), having a $G$ magnitude of 16.80 was detected with a period of only $\sim$\,23\,min by TMTS. Follow-up spectroscopic and photometric observations were taken and analyzed in 2024. This source is also reported by \cite{Jewett2024}, in a catalog paper focusing on massive WDs within 100\,pc. The physical parameters of J0006 they provide are $T_{\rm eff} = 25,442 \pm 522$\,K, $M=1.138 \pm 0.010$\,$M_{\odot}$, and cooling age = 0.20$\pm$0.01\,Gyr. More recently, \cite{Moss2025} estimate the magnetic field strength of J0006 to be 93\,MG based on the SDSS spectrum in their study of magnetic WDs in the SDSS 100\,pc sample. In this work, we present our new spectroscopic and photometric observations and a detailed analysis of J0006. The data are summarized in Section 2;  results and analysis are in Section 3. Section 4 presents
our discussion and conclusions.

\section{Observations}
A periodic signal of 23.09\,min was detected in the TMTS light curve of J0006, based on observations obtained on Sep. 23, 2022 (UTC; hereafter). The presence of this $\sim 23$\,min period is independently confirmed by both {\it TESS} and Zwicky Transient Facility (ZTF) light curves. Subsequently, follow-up photometry was carried out using the 0.8\,m telescope at Xinglong Observatory, China, and the 1.5\,m AZT-22 telescope at the Maidanak Astronomical Observatory in Uzbekistan \citep{Ehgamberdiev2018}. Spectra were obtained with the Kast double spectrograph \citep{Miller1993} on the Lick 3\,m Shane telescope and the Low-Resolution Imaging Spectrograph \citep[LRIS;][]{Oke1995} mounted on the Keck-I 10\,m telescope. The telescopes, their instruments, UTC observing dates, and exposure times are summarized in Table.\,\ref{tab1}. 

With the 2.4\,m telescope located at Lijiang Observatory, the Yunnan Faint-Object Spectrograph and Camera (YFOSC) was used to obtain the $g$-, $r$-, and $i$-band light curves of J0006. Each light curve in a given band covers about 1\,hr of continuous observations. Owing to less favorable weather conditions, the quality of these light curves is not as good as those obtained with AZT. The AZT data were obtained in the $B$, $V$, and $R$ bands, with each light curve spanning about 2\,hr. Our light curves are shown in Figure \ref{fig2}.

Spectra obtained with Lick/Kast consisted of three exposures of 1800\,s each, employing a 2$^{\arcsec}$-wide slit, the 600/4310 grism on the blue side (3634--5700\,AA), and the 300/7500 grism on the red side (5430--10,732\,\AA). 
We obtained 11 low-resolution Keck/LRIS spectra with individual exposure times of 240\,s, using a 1$^{\arcsec}$-wide slit, the 400/3400 blue grating, and the 400/8500 red grating. The resulting wavelength coverage is 3078--10,278\,\AA, with spectral dispersions of 1.07\,\AA/ pixel on the blue side and 1.19\,\AA/pixel on the red side.

To investigate possible X-ray emission from J0006, we  triggered a target-of-opportunity (ToO) observation with the Follow-up X-ray Telescope (FXT) onboard the {\it Einstein Probe (EP)} X-ray space telescope.

\begin{figure*}[!htbp]
\center
\includegraphics[angle=0,width=0.45\textwidth]{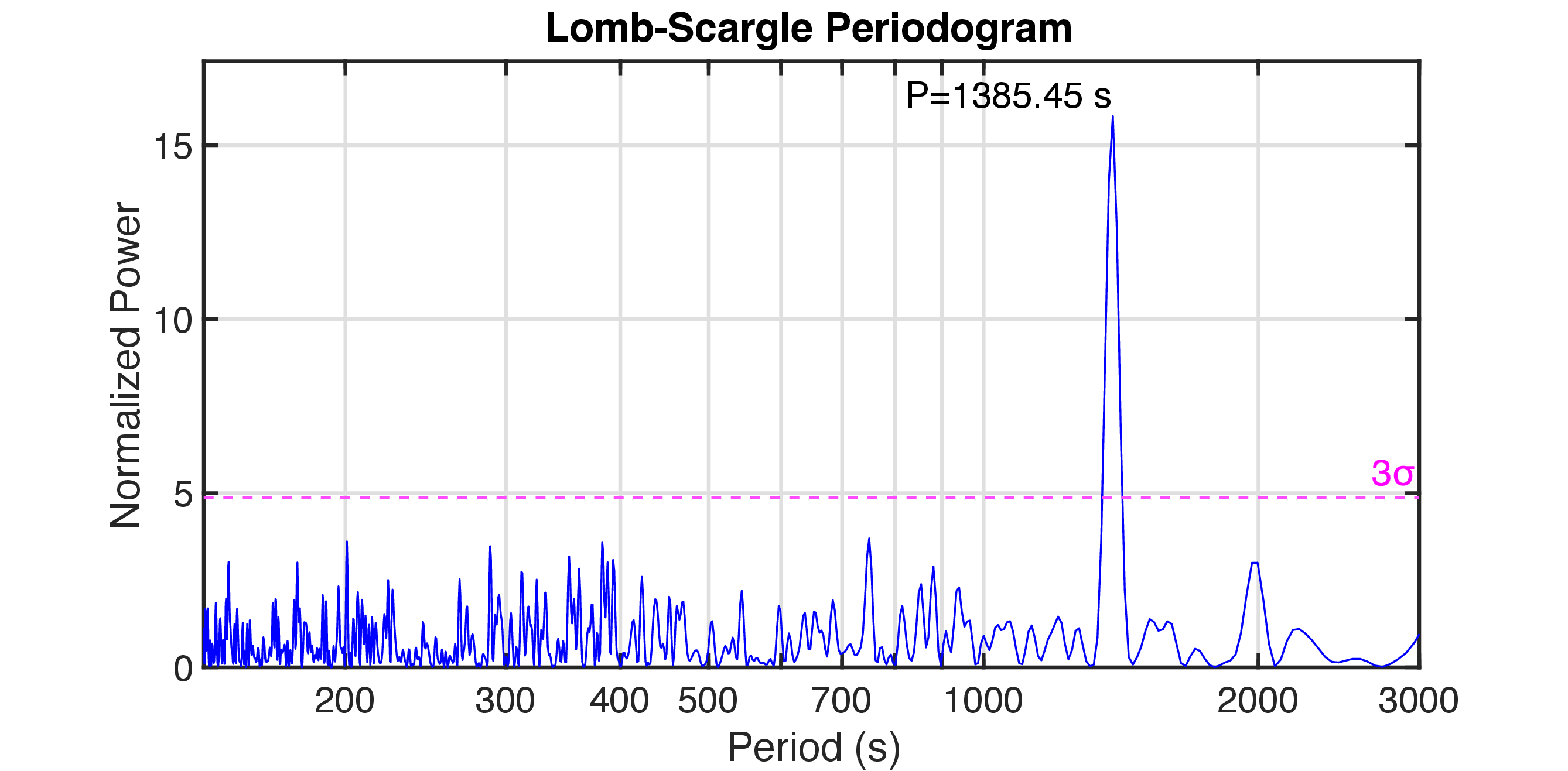}
\includegraphics[angle=0,width=0.44\textwidth]{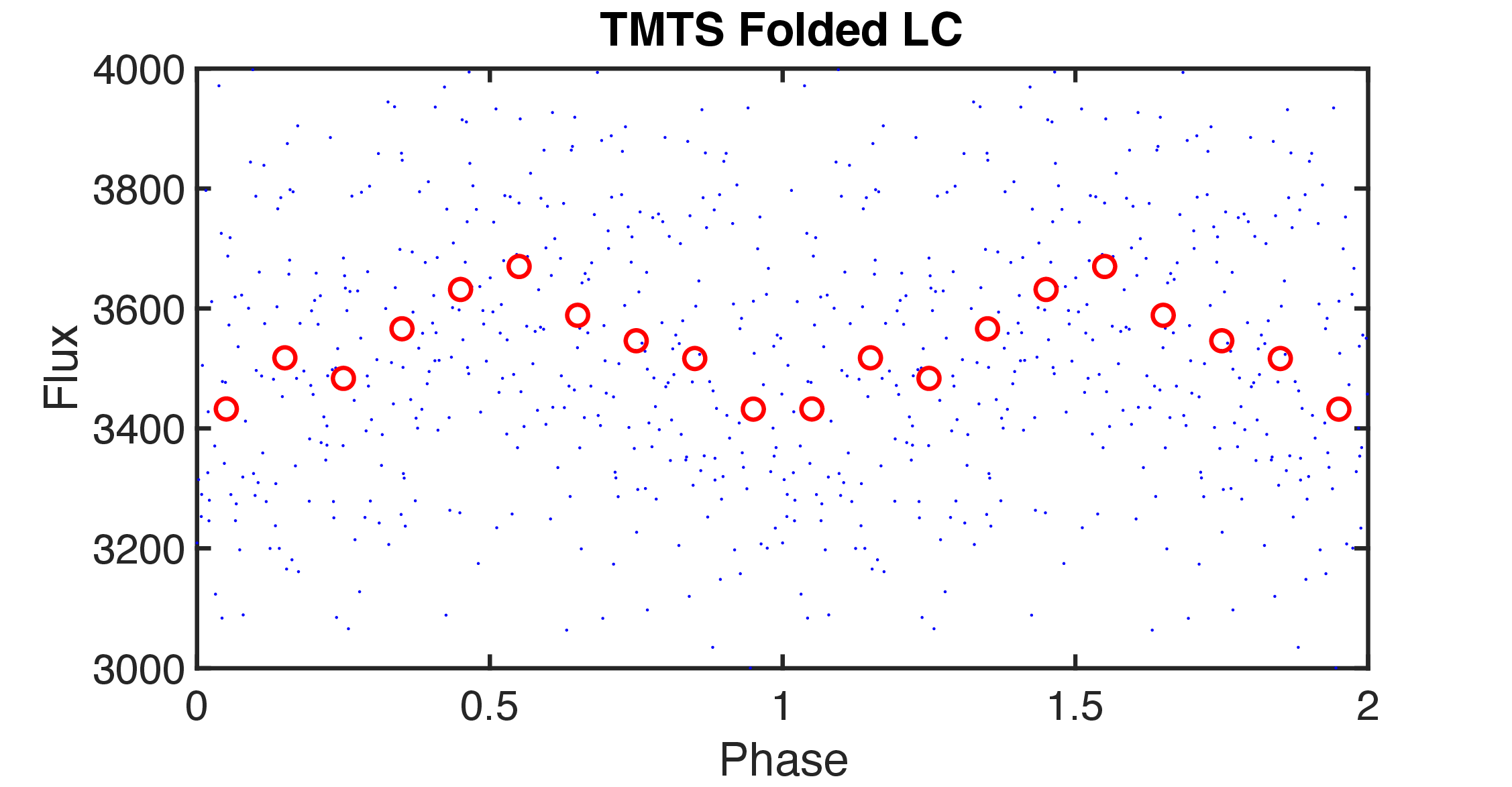}
\includegraphics[angle=0,width=0.45\textwidth]{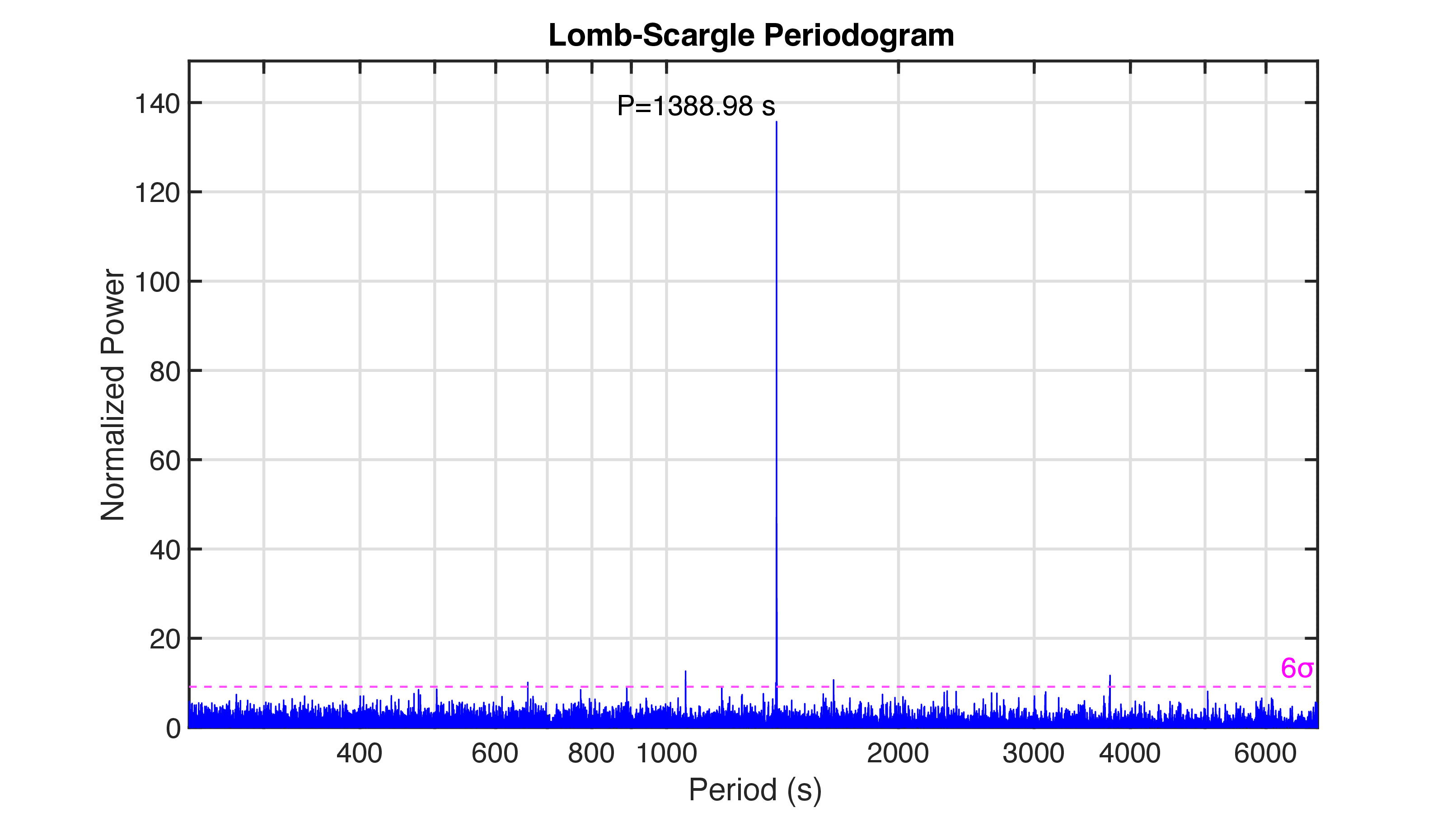}
\includegraphics[angle=0,width=0.45\textwidth]{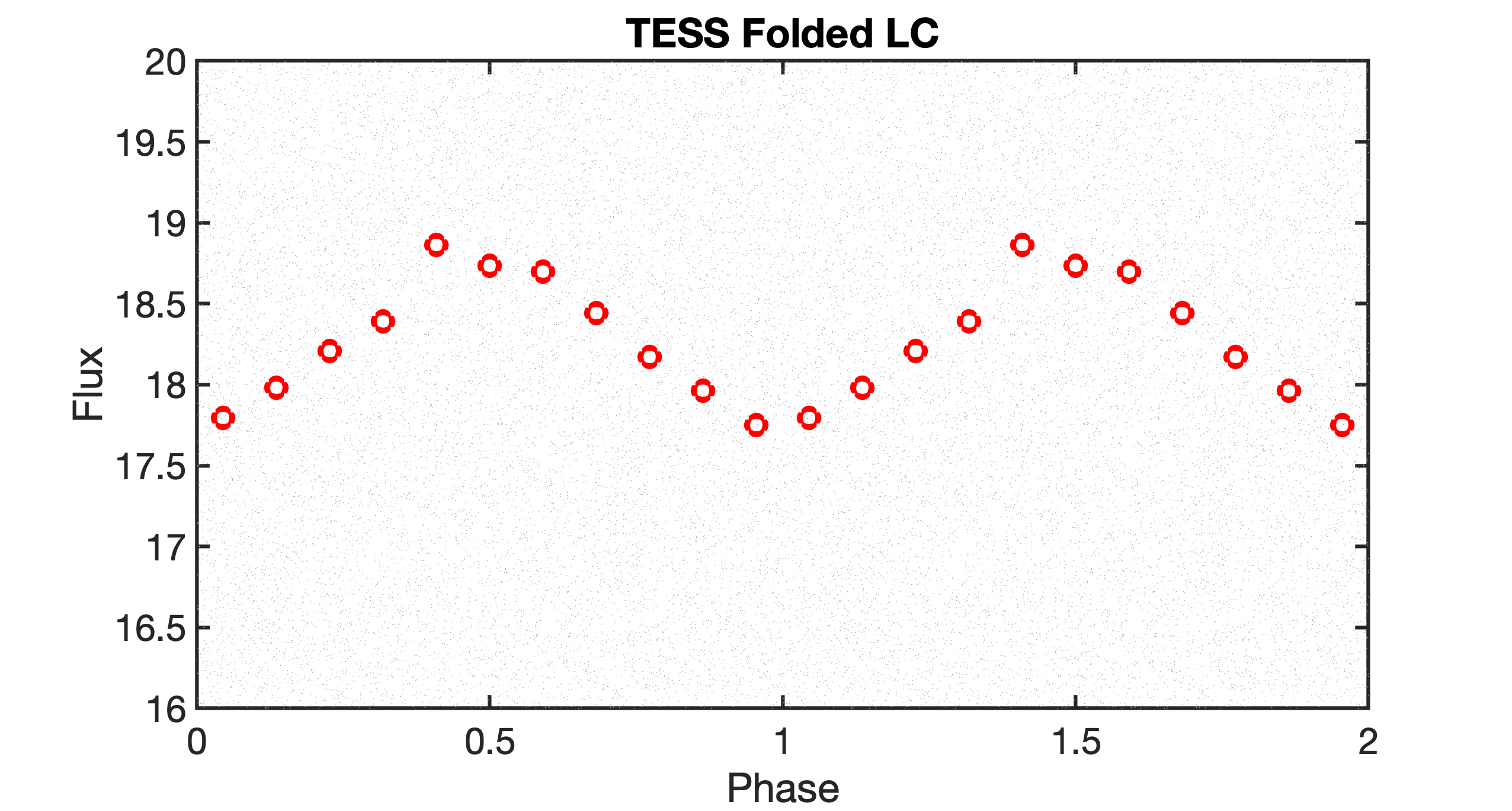}
\includegraphics[angle=0,width=0.45\textwidth]{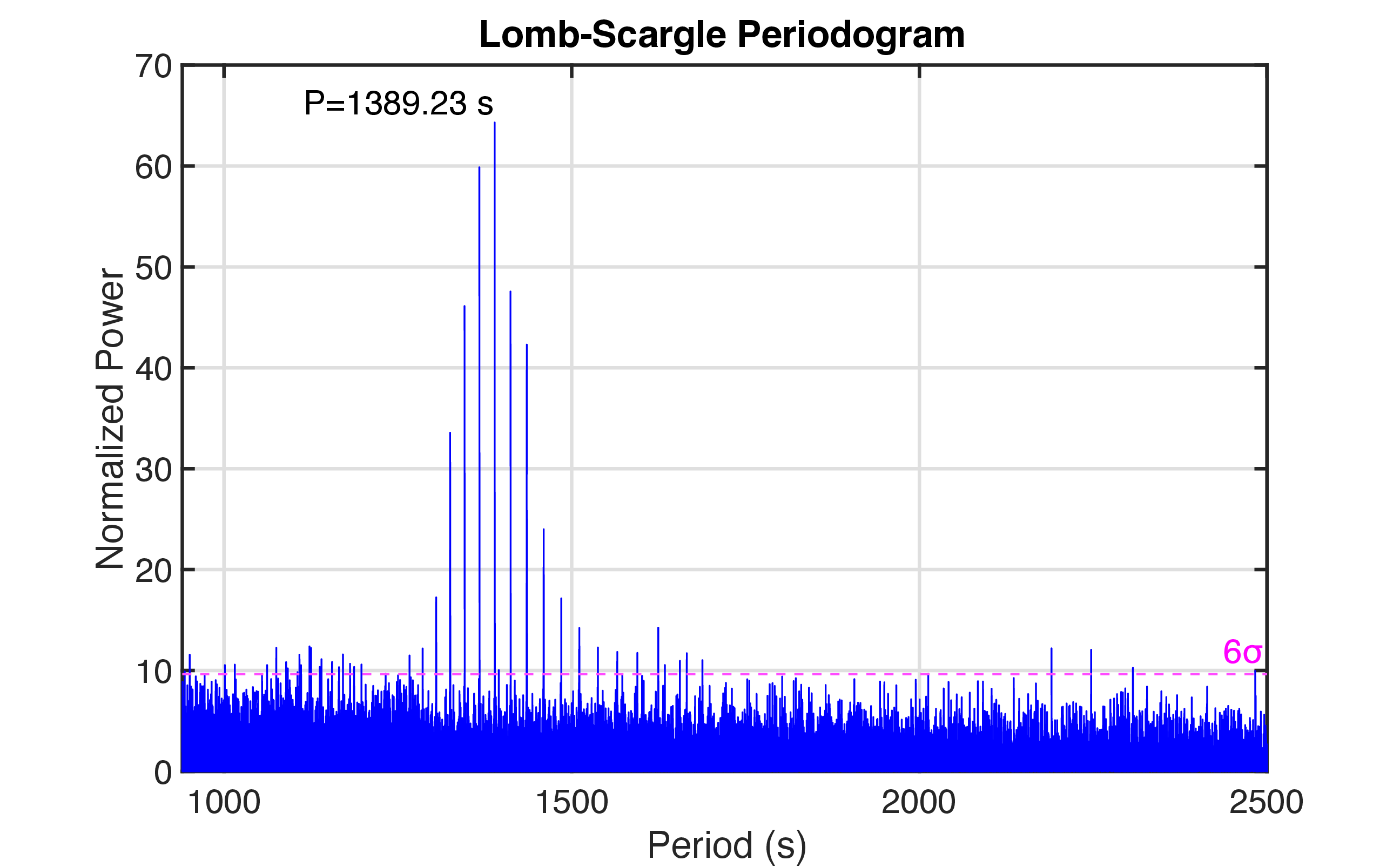}
\includegraphics[angle=0,width=0.45\textwidth]{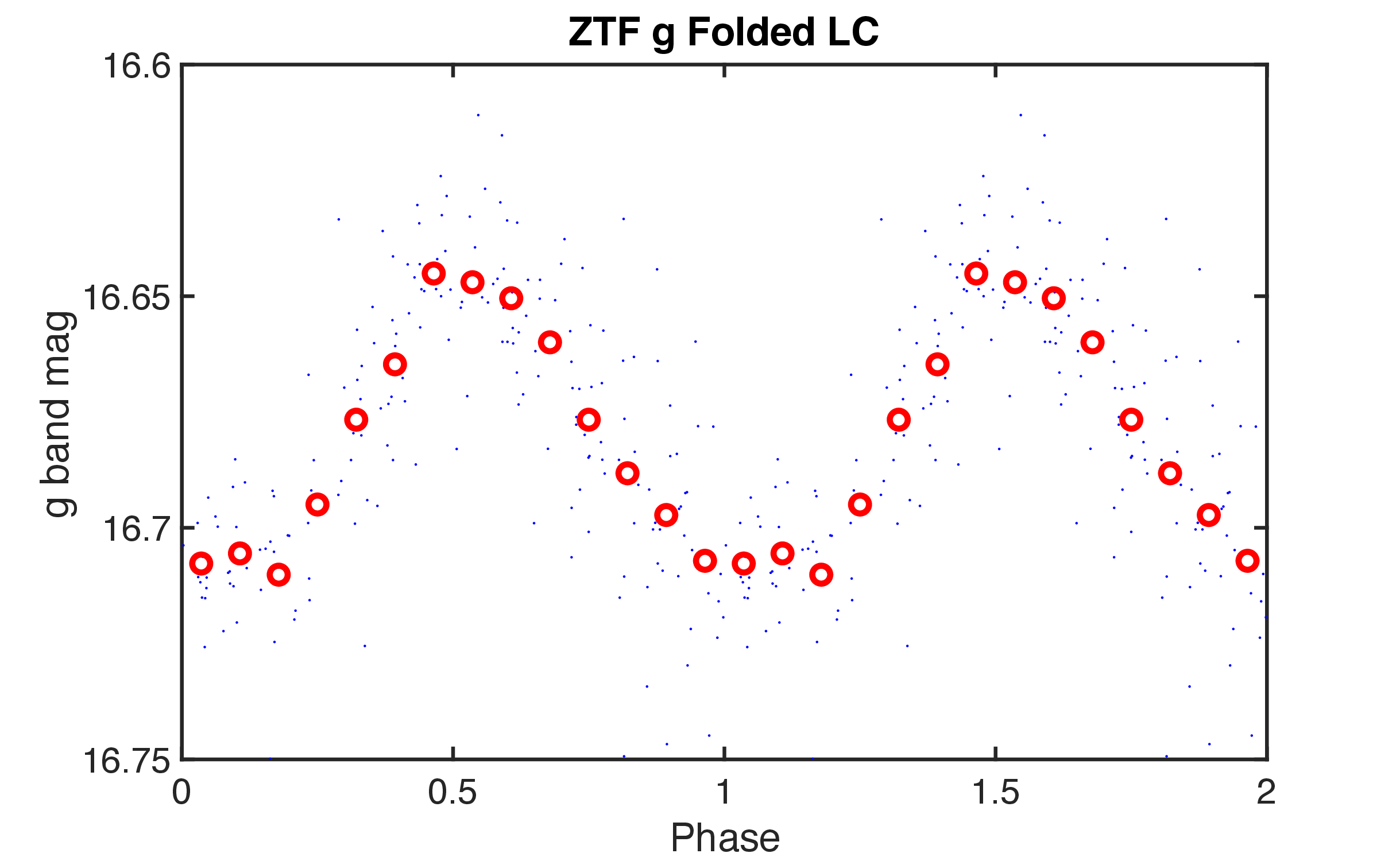}
\includegraphics[angle=0,width=0.45\textwidth]{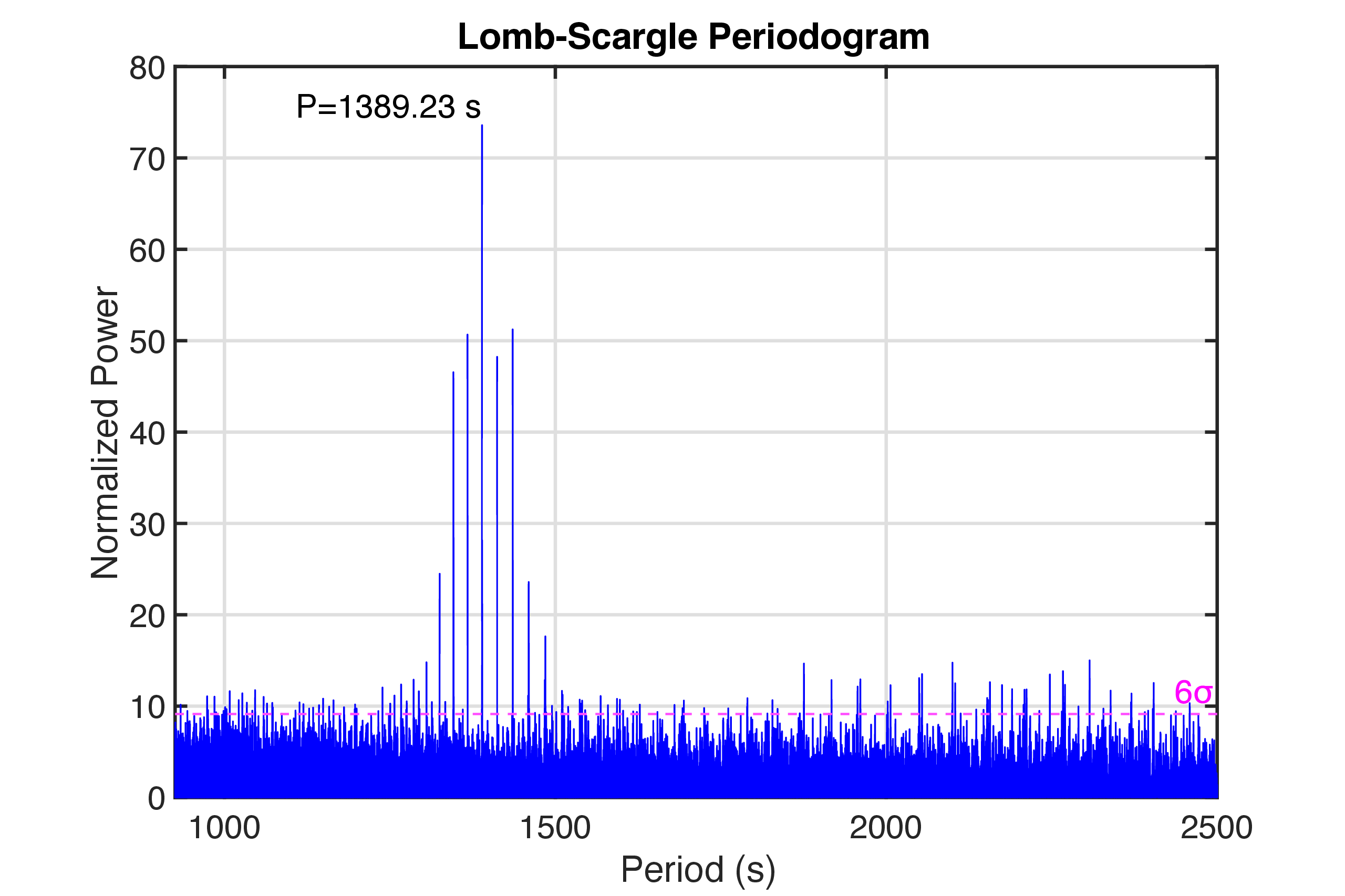}
\includegraphics[angle=0,width=0.47\textwidth]{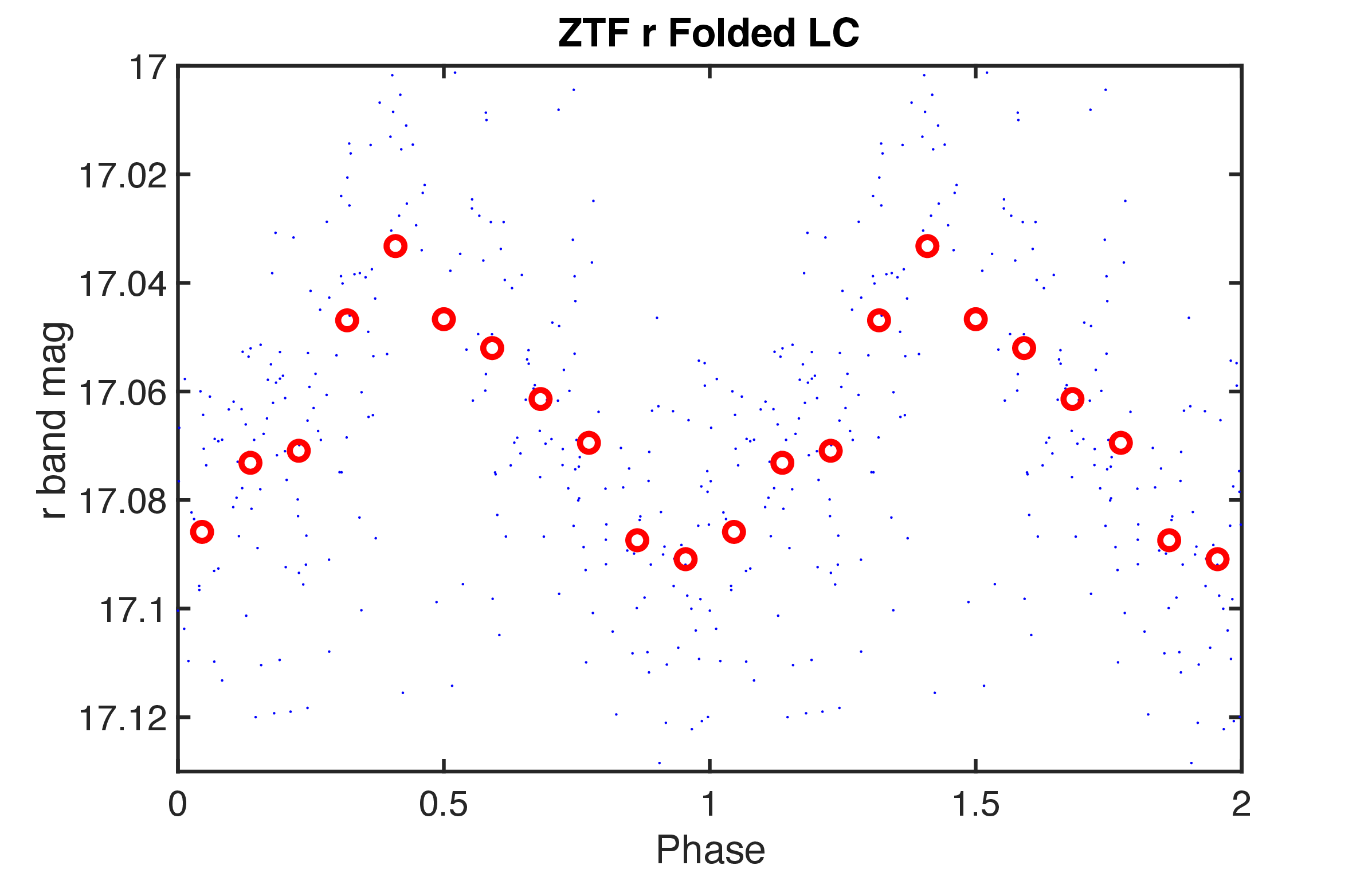}
\caption{Lomb-Scargle periodograms and folded light curves of TMTS, {\it TESS}, and ZTF $g$ and $r$ bands. In the left panels, the dashed magenta lines indicate the 3$\sigma$ or 6$\sigma$ significance levels, while the periods corresponding to the highest power are labeled. In the right panels, the phase-folded light curves are illustrated. The blue points are observed folded light-curve points. The red circles show the median flux or magnitude in phase bins to highlight the overall modulation. For display purposes, two phase cycles are plotted.}
\label{fig1}
\end{figure*}

\begin{figure*}[!htbp]
\center
\includegraphics[angle=0,width=0.9\textwidth]{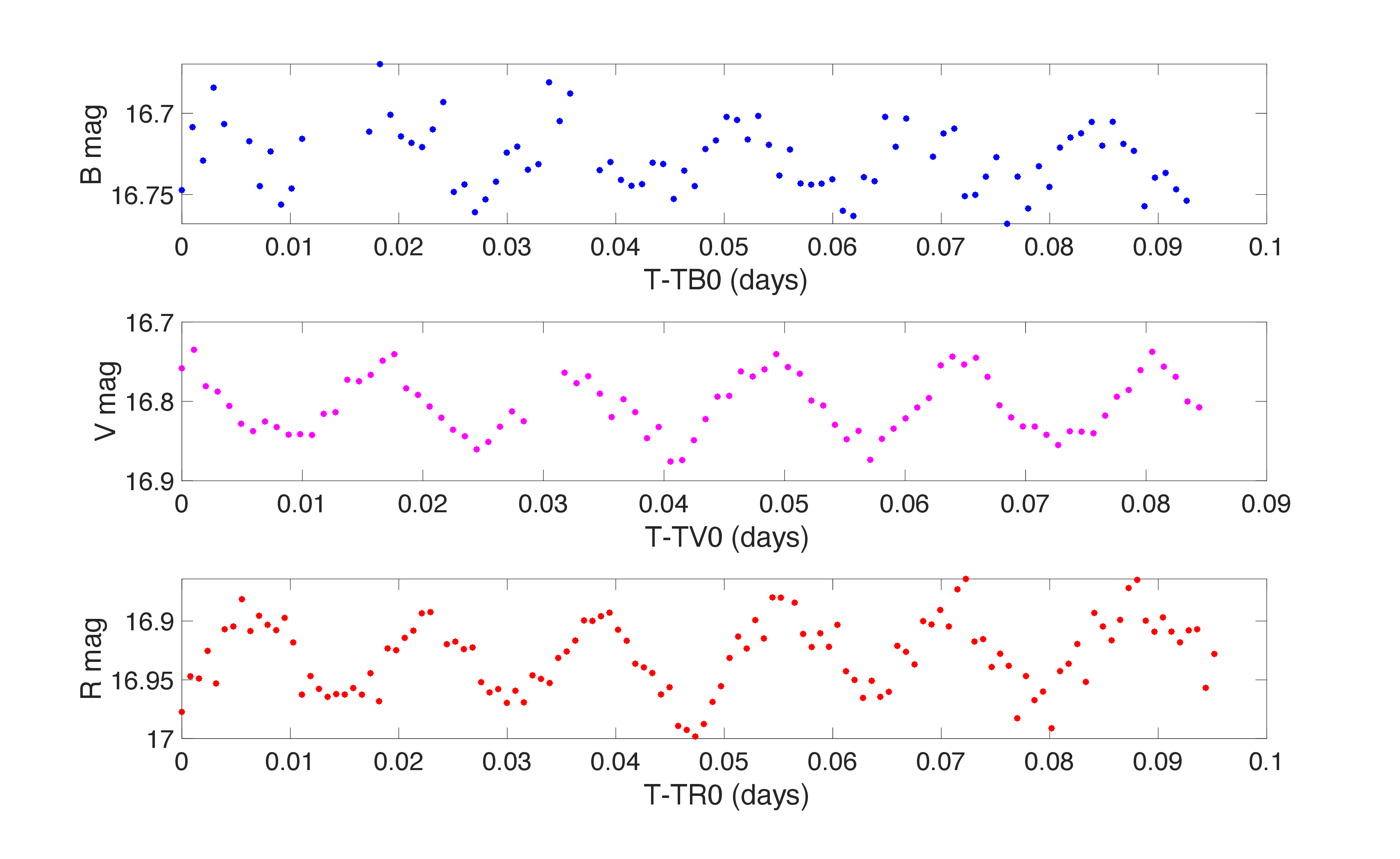}
\caption{Uninterrupted light curves obtained with the AZT 1.6\,m telescope in the $B$, $V$, and $R$ bands. The horizontal axes represent Heliocentric Julian Date (HJD) relative to the first observation epoch, while the vertical axes display the apparent magnitudes in the $B$, $V$, and $R$ bands. }
\label{fig2}
\end{figure*}

\begin{table}[]
\scriptsize
\begin{center}
\caption{Observations of J0006.}
\label{tab1}
\begin{tabular}{llll}
\\ \hline \hline
Telescope & Instrument & UTC Date & Exp. (s) \\   \hline
TMTS &  CMOS & 2022-09-23 & 60    \\
AZT-1.5\,m &  $B/V/R$  & 2023-12-03/20 &   60  \\
2.4\,m &  YFOSC($g/r/i$)  & 2023-09-16 &  45   \\
{\it TESS} & S17/S57/S84  & 2019;2022;2024& 20/120/1800 \\
ZTF & $g/r$  &   & 30   \\
0.8\,m -- IR &  $J$  & 2024-12-18 & 20\,s$\times$80    \\
\hline
Shane &  Kast  & 2023-11-10 & 1800\,s $\times$ 3        \\
Keck &  LRIS  & 2024-01-06 & 240\,s $\times$ 11    \\     \hline
{\it EP}   &  WXT   & 2024-12-28 & 1693             \\    
{\it EP}   &  FXT   & 2025-10-10 & 9533                   \\     \hline
\end{tabular}
\end{center}
\end{table}

\begin{figure*}[!htbp]
\center
\includegraphics[angle=0,width=0.47\textwidth]{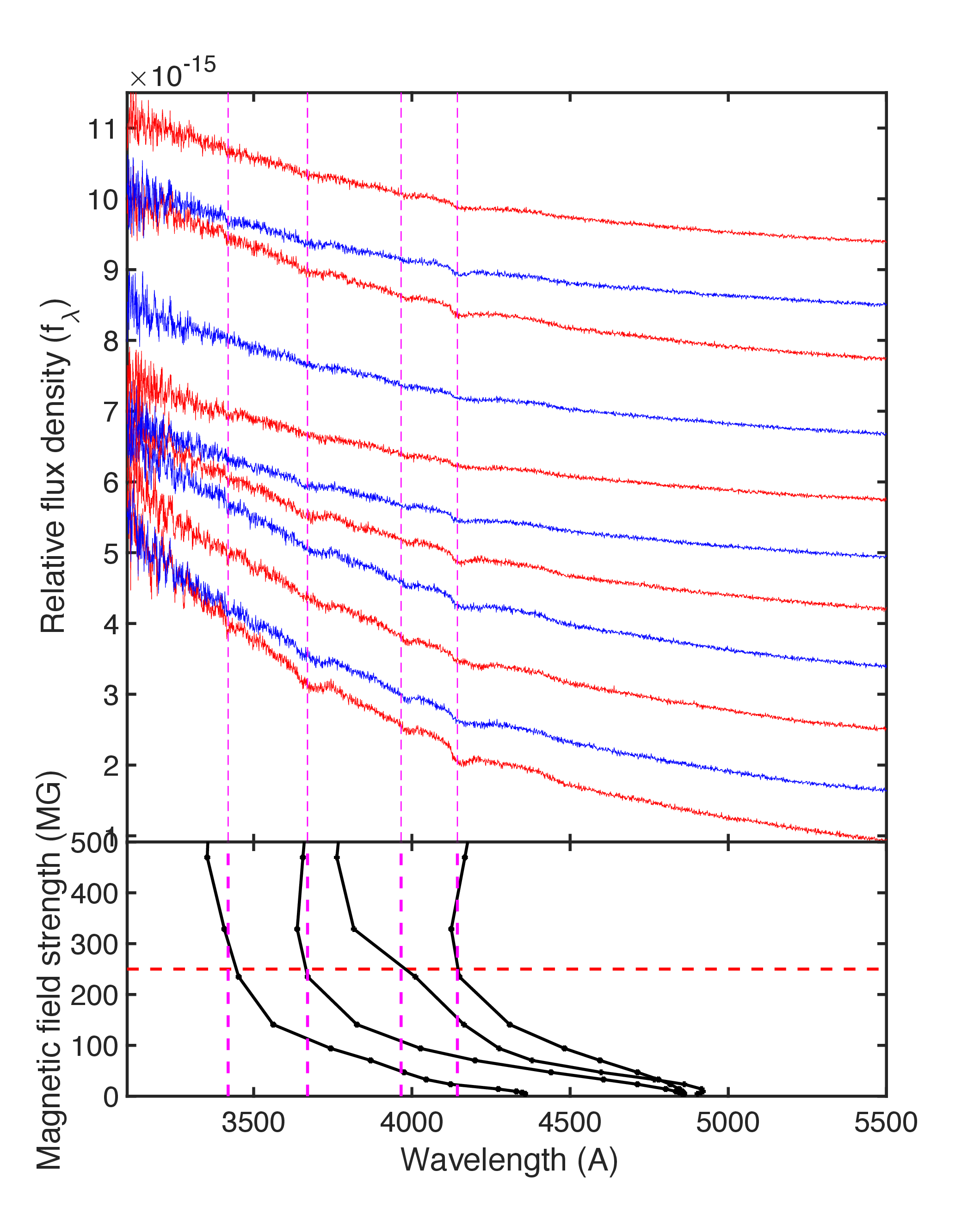}
\includegraphics[angle=0,width=0.47\textwidth]{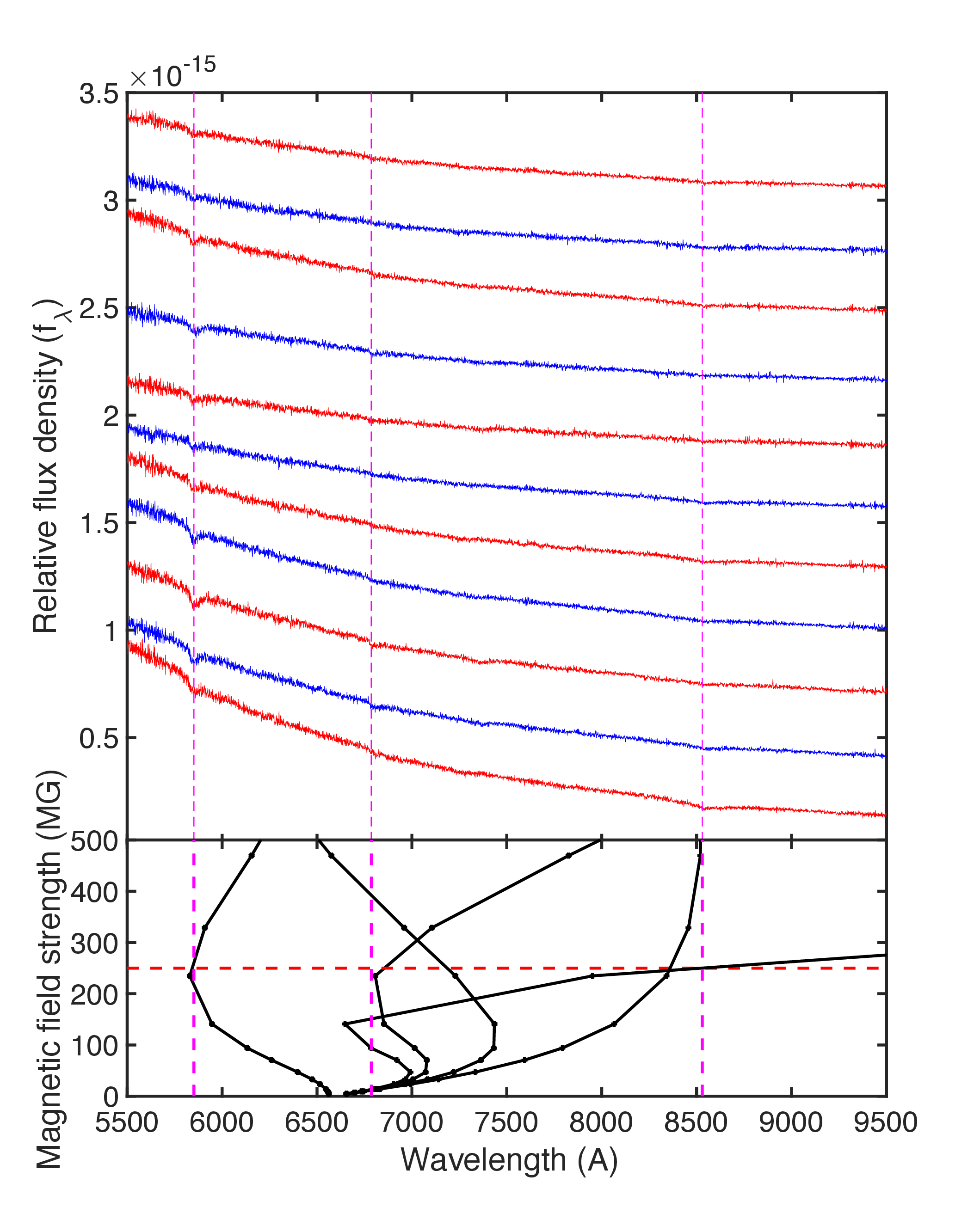}
\caption{Temporal series of Keck spectra of J0006. {\it Upper-left panel:} the blue side, with vertical offset. {\it Upper-right panel:} the red side. The {\it lower-left panel} corresponds to the magnetic field model to determine the magnetic field strength of the blue side, while the {\it lower-right panel} corresponds to the red side. Vertical dashed magenta lines mark the absorption lines, while the horizontal dashed lines display the 250\,MG strength.}
\label{fig3}
\end{figure*}

\section{Analysis and Results}
\subsection{Photometric Variation}
The short-period signal of $\sim 23$\,min is evident in the Lomb-Scargle (LS) periodograms of TMTS, {\it TESS}, and ZTF light curves, as shown in Figure \ref{fig1}. The specific periods identified are 1385.45\,s for TMTS, 1388.98\,s for {\it TESS}, and 1389.23\,s for ZTF, with all detected periods aligning within a 4\,s range. The {\it TESS} light curve comprises three sectors with 120\,s exposure time. Owing to the long duration, the highest significance in the LS periodogram, and being free from atmospheric influence, we consider the period of 1388.98\,s (23.15\,min) from {\it TESS} to be the most accurate. In addition, to estimate the ephemerides, we used the Sector 20 data obtained in 2019 with a 120\,s exposure to calculate $T_{0}$. Analysis of this light curve yields the following best-fit ephemerides: BJD$_{\rm TDB}=2458778.861(1)\,+\, 0.0160777(1)\, E$. 
Based on the phase-folded light curves, the shape of J0006's light curves does not indicate a binary system. The variation between peaks and minima in the TMTS data is $\sim 5$\%, while the variations in the ZTF $g$ and $r$ bands are roughly 0.06\,mag and 0.08\,mag, respectively. 

Figure \ref{fig2} presents the uninterrupted AZT light curves in the $B$, $V$, and $R$ bands. The $V$ and $R$  light curves were collected under better conditions, exhibiting symmetric variations with comparable peaks and minima. Their shapes align closely with those observed in the TMTS, {\it TESS}, and ZTF light curves. 

\subsection{Spectroscopy}
Following the discovery of J0006's short period, three exploratory spectra were taken with Lick/Kast in Dec. 2023, aimed at further characterizing this WD. All three exhibit a WD-like continuum (bright on the blue side, gradually dimming on the red side), accompanied by a few shallow but broad absorption features. Since the exposure time exceeded the period and the spectra had a low signal-to-noise ratio (S/N),  additional spectra were collected with Keck/LRIS in Jan. 2024. Figure \ref{fig3} illustrates these spectra, presented with offsets in relative flux density for better visualization. They  reveal more distinct line features that closely resemble those found in the spectra of the WD ZTF J190132.9+145808.7, as depicted in Fig. 3 of \cite{Caiazzo2021}. Despite this similarity, no significant line shifts were observed. 
Based on the presence of broad and shallow absorption features in the optical spectra, we compared the observations with hydrogen-dominated magnetic WD atmosphere model spanning a range of field strengths. We find that models with a magnetic field strength of approximately 250 MG reproduce the characteristic Zeeman-smeared absorption features remarkably well. In contrast, helium-atmosphere models fail to match the observed lines. We therefore classify J0006 as a hydrogen-atmosphere magnetic WD (DAH). This classification supersedes the previous DC and DB classifications reported by \citep{Jewett2024} and \citep{Vincent2024}, which likely arose from extreme magnetic broadening that masks diagnostic spectral features in lower-quality data.

\begin{figure}[!htbp]
\center
\includegraphics[angle=0,width=0.5\textwidth]{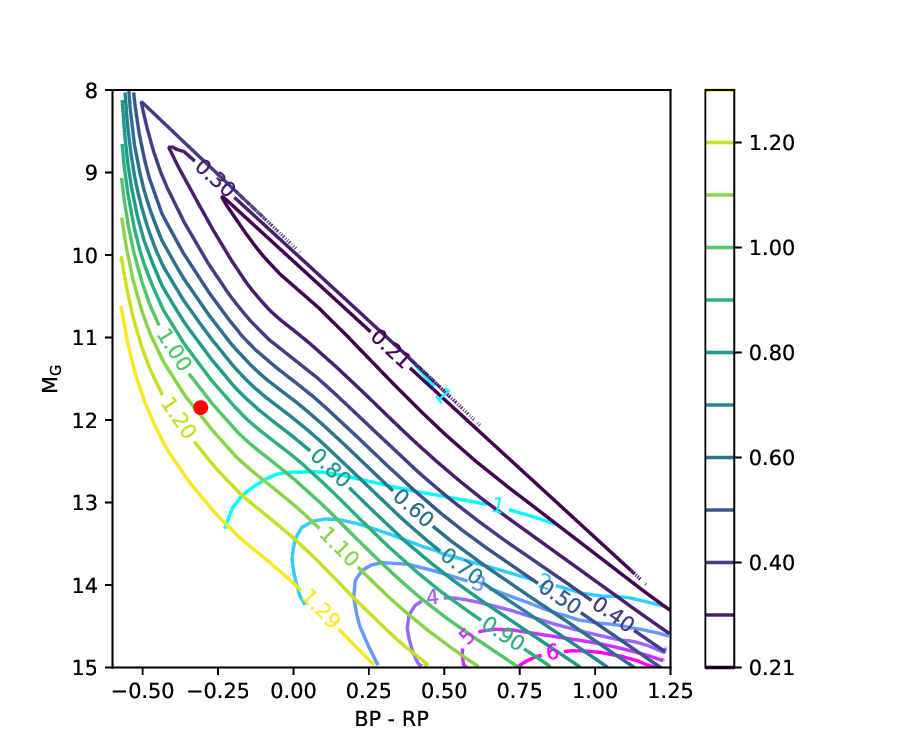}
\caption{Mass and cooling-age contours on the {\it Gaia} color vs. absolute-magnitude diagram. The H-rich DA model from \cite{Bedard2020} is adopted. The red dot marks the location of J0006. The contours from upper right to lower left are WD masses from 0.21\,M$_{\odot}$ to 1.29\,M$_{\odot}$. The contours from middle left to lower right are cooling ages from 1\,Gyr to 6\,Gyr.}
\label{fig4}
\end{figure}

\subsection{Gaia data}
Using precise astrometry and photometry from {\it Gaia} DR3, we determine atmospheric parameters for J0006. The parallax-derived distance of 98.103\,pc, BP--RP of $-0.3075$\,mag, and $G$-band magnitude of 16.808 place J0006 in the {\it Gaia} color-magnitude diagram above the 1\,M$_{\odot}$ locus (Figure\,\ref{fig4}). According to the WD model from \cite{Bedard2020}, parameters of J0006 are derived from {\it Gaia} data as follows: $T_{\rm eff} = 22,000 \pm 600$\,K, log\,$g = 8.73 \pm 0.02$ (where $g$ is in cgs units), $M = 1.06 \pm 0.01$\,M$_{\odot}$, and cooling age = $0.237\pm0.009$\,Gyr. Independently, parameters were estimated by fitting the low-resolution {\it Gaia} XP spectra \citep{Vincent2024}. They suggest that the spectral type of J0006 is ``DB:'' and the model-atmosphere composition is pure helium; parameters of J0006 are $T_{\rm eff} = 28,135 \pm 1,319$\,K, log\,$g = 8.912 \pm 0.018$, and $M=1.152\pm0.010$\,M$_{\odot}$.

\subsection{Spectral Energy Distribution Fitting}

\begin{figure}[!htbp]
\center
\includegraphics[angle=0,width=0.55\textwidth]{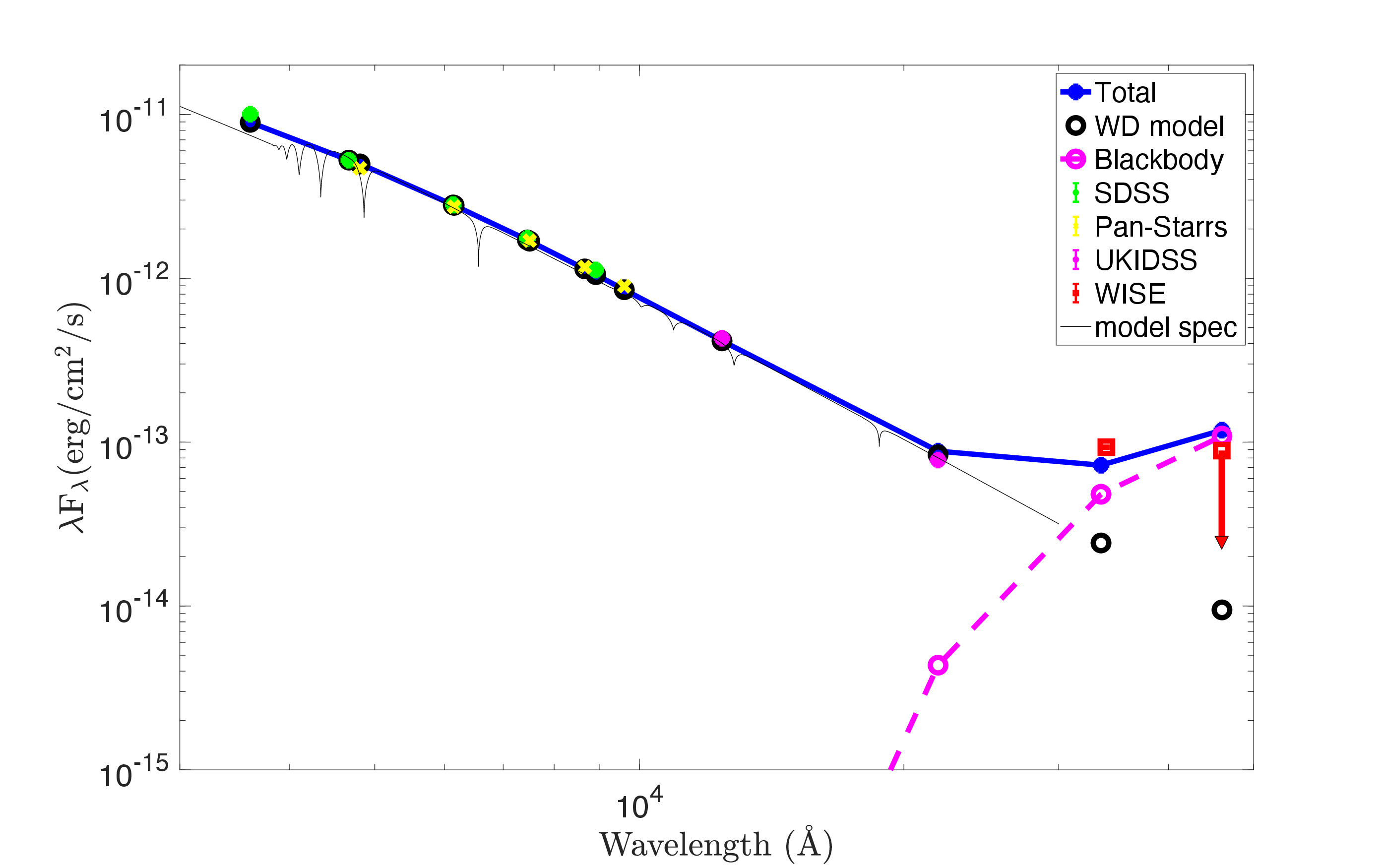}
\caption{SED fitting for J0006. The green dots represent data from SDSS in the $u$, $g$, $r$, $i$, and  $z$ bands. Yellow crosses indicate data from Pan-Starrs in the $g$, $r$, $i$, $z$, and $y$ bands. The best-fit model spectrum corresponding to 25,000\,K is overplotted. }
\label{fig5}
\end{figure}

 To further constrain the physical properties of J0006, we constructed its spectral energy distribution (SED) using broad-band photometry from SDSS, Pan-Starrs, UKIDSS UHS (the UKIRT Hemisphere Survey), and {\it WISE}. At present, publicly available white dwarf atmosphere models that simultaneously account for extreme magnetic fields and detailed broadband flux distributions remain limited. We therefore adopt DA WD model spectra from \cite{Koester2010} as a practical approximation for the purpose of broadband SED fitting. We emphasize that these models are used only to estimate basic atmospheric parameters (e.g., $T_{\rm eff}$ and radius), while the magnetic field strength and spectral classification are determined independently from spectroscopic analysis. The best-fitting solution yields an effective temperature of 25,000\,K and negligible extinction ($A_{V}=0$\,mag). Furthermore, the radius of J0006 ($R_{\star}$) can be estimated to be 0.71\,R$_{\oplus}$ using the relation $F_{\lambda}^{obs}/F_{\lambda}^{model}=4\pi\,R_{\star}^{2}/4\pi\,D^{2}$. The proportionality (flux-scaling or normalization) factor $M_{d}$ is defined as $M_{d}=F_{\lambda}^{obs}/F_{\lambda}^{model}$. Thus, from SED fitting $M_{d}=(R_{\star}/D)^2=2.313\times10^{-24}$, where $D=98.103$\,pc is the distance to the WD from Gaia. More notably, a clear infrared (IR) excess is detected in the $W1$ band of {\it WISE}. This excess cannot be reproduced by the WD photospheric model and is well fitted by an additional blackbody component. The best-fitting temperature of this component is approximately 550\,K, shown as the magenta curve in Figure \ref{fig5}. We note that uncertainties associated with the use of non-magnetic atmosphere models do not affect the detection or characterization of the IR excess, which is robust against reasonable variations in the photospheric model. The presence of this cool IR component therefore provides independent evidence for circumstellar material associated with J0006.

\section{Discussion}

\subsection{Origin of the Magnetic Field}
The existence of a magnetic field in WDs is common, usually accompanied by  Zeeman splitting in their spectra. However, WDs with magnetic field strength exceeding 100\,MG are rare and complicated. In these extreme cases, the absorption lines are strongly shifted and distorted, making an accurate estimation of the field strength challenging.

Based on the features of J0006, namely its short rotation period, high mass, and strong magnetic field, this source is most likely  the merger remnant of two WDs. It is widely accepted that a strong magnetic dynamo could arise from the merger process. More specifically, a hot, differentially rotating, convective corona generated from the merger of two degenerate cores is able to produce a strong magnetic field, which could be confined to the outer layers of the merger remnant and persist over long timescales \citep{Garcia-Berro2012}. Recent observations support this scenario for high-mass magnetic WDs, showing mixed polarimetric variability consistent with diverse magnetic geometries produced during the merger dynamo \citep{Bagnulo2024}. In this scenario, the ultrashort rotational period can be understood as well; the conservation of binary WD orbital angular momentum would easily result in the rapid rotation of the merger remnant. On the other hand, the magnetic field of an isolated WD could also be inherited from the previous evolution of its progenitor \citep{Angel1981}. Recent studies suggest an alternative or complementary fossil-field channel, where magnetic fields generated by convective-core dynamos during the main-sequence phase of more massive progenitors ($>\sim1.1-2\,M_{\odot}$) survive internal disruptions and gradually diffuse to the surface during WD cooling, emerging earlier and stronger in higher-mass objects \citep{Camisassa2024,Castro2025}. But this case cannot explain the ultrashort period and large mass observed in J0006. 

The third possibility is that strong magnetic fields are generated when a WD binary evolves through the common-envelope phase, where the companion of the WD caused differential rotation in the envelope while spiraling in. Nevertheless, magnetic fields produced in this manner would decay quickly after the ejection of the common envelope \citep{Potter2010}.

\subsection{Source of Variability}
Given that the effective temperature of J0006 ($22,000 \pm 600$\,K) lies well beyond the blue edge of the DA variable (DAV or ZZ Ceti stars) instability strip, the observed period should not be explained by DAV pulsation. While the temperature of J0006 falls within the range of DBVs, which is 22,400\,K to 32,000\,K \citep{Corsico2019}, even the maximum known pulsation period of DBVs (1080\,s) is shorter than the period detected in J0006. Furthermore, the absence of multiple significant frequencies in the LS periodograms disfavors a pulsational origin of the variability. We also consider an orbital interpretation to be unlikely, as no significant radial-velocity variations are detected in our time-series spectra. Taking into account the strong magnetic field of J0006, we thus attribute the observed periodicity to the rotation of the WD, potentially modulated by surface spots or occultation by circumstellar material.

\subsection{Infrared Excess Caused by Merger}
The new feature we found in J0006 among its kind is the presence of a significant  infrared (IR) excess detected in the {\it WISE} observations. By fitting the SED constructed from SDSS, Pan-Starrs, UKIDSS UHS, and {\it WISE} photometry, the best-fitting WD model yields $T_{\rm eff} = 25,000$\,K. A significant excess is clearly detected in the {\it WISE} $W1$ band, which can be well fitted by a 550\,K blackbody. The possibility of an unseen brown dwarf (BD) companion is investigated by fitting combined WD and BD models \citep{Phillips2020}. A 400\,K BD model is found with a very poor fit. Possible explanations include an unseen companion and a debris disk formed through the tidal disruption of an asteroid. However, it is more plausible that the IR excess is evidence of a double WD merger, arising from residual material associated with the merger remnant.

\subsection{Galactic Population}
Owing to the fact that the radial velocity of J0006 cannot be reliably measured from its highly magnetic spectra, the Galactic population of J0006 can instead be roughly constrained by estimating its tangential velocity. Based on the {\it Gaia} proper motions of PMRA=21.653\,mas\,yr$^{-1}$ and PMDEC=$-25.702$\,mas\,yr$^{-1}$, together with a distance of 98.103\,pc derived from the {\it Gaia} parallax, we obtain a tangential velocity of 15.63\,km\,s$^{-1}$. This relatively low velocity strongly suggests that J0006 belongs to the thin-disk population, as typical tangential velocities for thin-disk stars are on the order of a few tens of km\,s$^{-1}$.

\subsection{Nondetection of X-rays}
\begin{figure}[!htbp]
\center
\includegraphics[angle=0,width=0.5\textwidth]{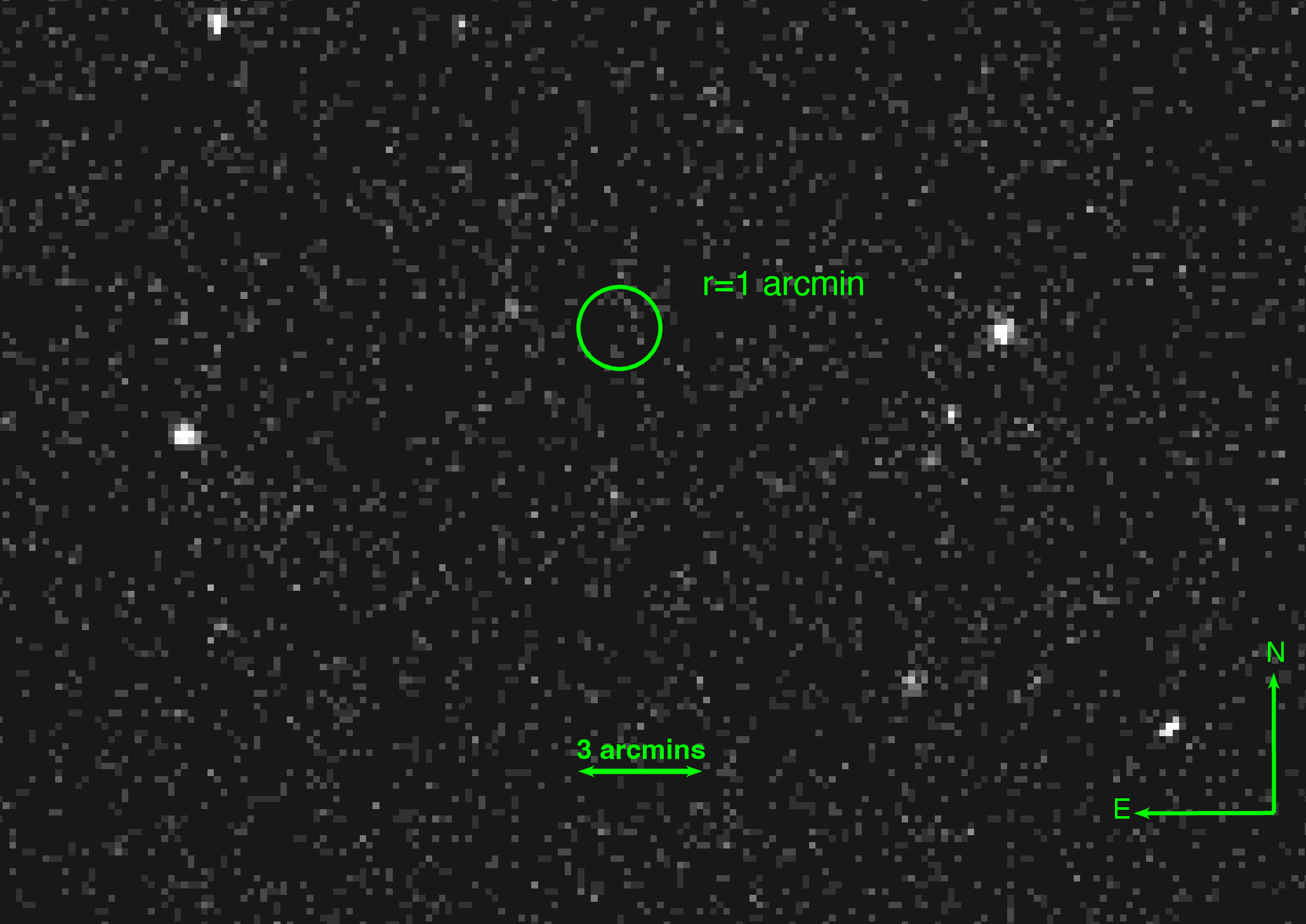}
\caption{FXT follow-up observation of J0006. Green circle marks the position of J0006 with a radius of 1$^{\prime}$.}
\label{fig51}
\end{figure}

For the purpose of exploring  possible X-ray emission from J0006 associated with the strong magnetic field, we proposed a ToO observation with the FXT onboard the {\it EP} with a total exposure time of 9533\,s. After standard data reduction and analysis using \texttt{Xselect}, no point source is detected within a  1$^{\prime}$ radius. A 5$\sigma$ upper limit of $1.26\times10^{-13}$\,erg\,s$^{-1}$\,cm$^{-2}$ in the 0.5--10.0\,keV band is estimated (see Fig.\ref{fig51}). According to \citet{Cristea2025}, the massive WD merger remnant ZTF J2008+4449 with a 6.6\,min rotational period and a period derivative of $(1.80\pm0.09)\times10^{-12}$\,s\,s$^{-1}$ is clearly detected in X-rays by {\it XMM-Newton}. Their X-ray emission is explained by the fallback of gravitationally bound material from the merger or from the tidal disruption of a planetary object. By comparison, a rough upper limit on the timescale for the disappearance of X-ray emission is $\sim 17$\,Myr  ($5.52\times10^{14}$\,s), based on the increase of the rotational period from 6.6\,min to 23.15\,min.

\subsection{Comparison with Other Similar WDs}
\begin{figure}[!htbp]
\center
\includegraphics[angle=0,width=0.55\textwidth]{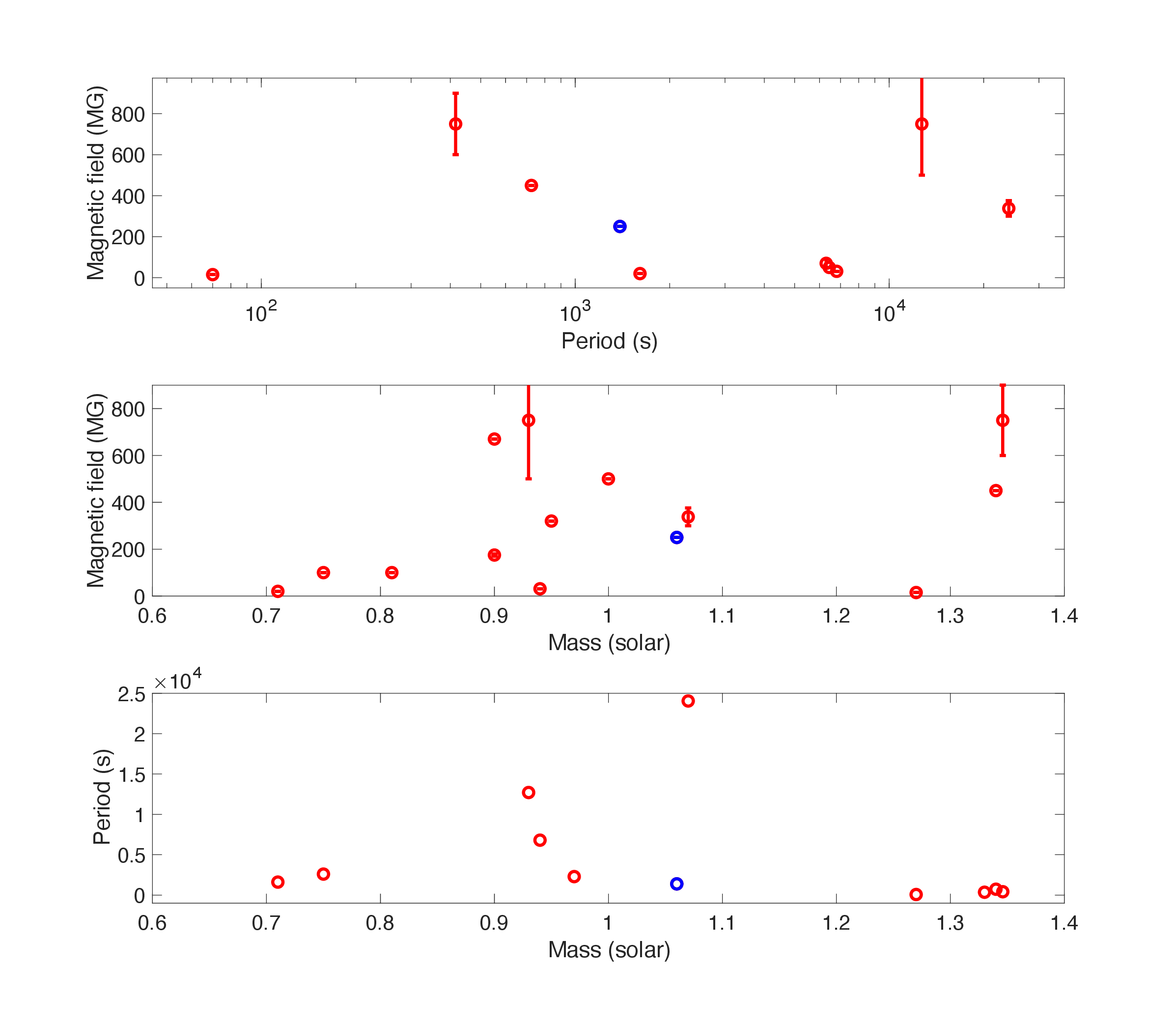}
\caption{Comparisons of period, mass, and magnetic field of a selected WD sample with short periods and strong magnetic fields. From top to bottom: period vs. magnetic field, mass vs. magnetic field, and mass vs. period. Red circles are similar WDs from the literature, while the blue circle represents J0006.}
\label{fig6}
\end{figure}

To date, about a dozen WDs with similar properties (i.e., short rotation periods, strong magnetic fields, large masses) have been discovered. It is worthwhile to study both their common and distinct characteristics. In Figure \ref{fig6}, we compile WDs with either short periods or strong magnetic fields from the literature \citep{Barstow1995,Brinkworth2013,Reding2020,Caiazzo2021,Kilic2021,Williams2022,Yan2025,Cristea2025}. Comparisons are made between their period and magnetic field, mass and magnetic field, and mass and period. Within the period range of 100 to 2000\,s, the magnetic field strength appears to increase as the period decreases, although the current sample size is still insufficient for a firm statistical conclusion. 

For the comparison of mass and magnetic field, massive WDs ($>0.9$\,M$_{\odot}$) tend to have much stronger magnetic fields, indicating a merger origin for high-mass and strongly magnetic WDs. Only two objects deviate from this trend, showing high mass but relatively weak magnetic fields; these may have originated from single massive progenitors or represent merger remnants whose magnetic fields have decayed over long timescales. For the comparison between mass and period, the majority of objects in our selected sample are detected with short period ($<$2\,hr). Especially when the masses are around 1.3\,M$_{\odot}$, their corresponding periods are $<12$\,min, suggesting a merger origin most likely; ultrashort periods are generated from the conservation of orbital momentum.

\section{Summary}

In this work, we present the first comprehensive physical characterization of the ultra massive, highly magnetic, fast-rotating WD J0006, identified by TMTS. Photometric observations reveal a period of 23.15\,min. Low-resolution Keck spectroscopy indicates the presence of a strong magnetic field, with no detectable line shifts. According to the adopted magnetic models, the spectral type of J0006 is DAH and field strength is estimated to be $\sim 250$\,MG. Based on accurate astrometry from the {\it Gaia} DR3, J0006 is located at a distance of 98\,pc. With the help of {\it Gaia} photometry and parallax, as well as a WD atmosphere model, its parameters are determined to be $T_{\rm eff} = 22,000 \pm 600$\,K, log\,$g = 8.73 \pm 0.02$ (where $g$ is in cgs units), $M = 1.06 \pm 0.01$\,M$_{\odot}$, and cooling age $ = 0.237 \pm 0.009$\,Gyr. From the SED fitting, the radius of J0006 is estimated to be 0.71\,R$_{\oplus}$. 

More interestingly, a significant IR excess is detected in the {\it WISE} $W1$ band. A 400\,K BD companion model is found with a very poor fit, whereas a 550\,K blackbody can well fit the excess. This suggests that the IR excess is more likely associated with residual material from a binary merger. 

Through comparison with other highly magnetic and short-period WDs, no definite correlations are found, mainly limited by the small sample size. Nevertheless, the available data indicate that most WDs with masses greater than 0.9\,M$_{\odot}$ tend to host strong magnetic fields ($> 200$\,MG), while those with masses around 1.3\,M$_{\odot}$ preferentially exhibit very short periods ($<2$\,hr). With a rotational period of only 23.15\,min, J0006 is among the shortest-period double WD merger remnants known, exhibiting a magnetic field strength of hundreds of MG and no detectable X-ray emission. 

The key properties of J0006 identified in this work—namely, the presence of a 550 K blackbody infrared excess, the non-detection of X-ray emission, the ultra-short rotation period, and a magnetic field strength of several hundred MG—collectively suggest that it represents an intermediate evolutionary stage between ZTF J2008+4449 \citep{Cristea2025} and more typical highly magnetic, rapidly rotating white dwarfs from binary merger. By comparison, we very roughly estimate that ZTF J2008+4449 could evolve into a system similar to J0006 after $\sim$17 Myr, characterized by a longer rotation period, the disappearance of X-ray emission, and a cooler temperature of the merger residual material.

\section*{Acknowledgments}
The authors acknowledge the National Natural Science Foundation of China (NSFC) under grants 12288102, 12203006, 12033003, 12273055, 12503027, and 12422305; the Young Scholar Program of Beijing Academy of Science and Technology (25CE-YS-02,24CE-YS-08), the Tencent Xplorer Prize, the Ma Huateng Foundation, and the Beijing Natural Science Foundation (1242016). J.C.G. and C.L. acknowledge helpful discussions with Prof. Xianfei Zhang. The team from Uzbekistan carried out this work at the Ulugh Beg Astronomical Institute of the Academy of Sciences of Uzbekistan, within the research program of the Advanced Astrophysical Research Laboratory.

A.V.F.'s research group at U.C. Berkeley received financial assistance from the Christopher R. Redlich Fund, as well as donations from Gary and Cynthia Bengier, Clark and Sharon Winslow, Alan Eustace and Kathy Kwan, William Draper, Timothy and Melissa Draper, Briggs and Kathleen Wood, Sanford Robertson (W.Z. is a Bengier-Winslow-Eustace Specialist in Astronomy, T.G.B. is a Draper-Wood-Robertson Specialist in Astronomy, 
Y.Y. was a Bengier-Winslow-Robertson Fellow in Astronomy), 
and numerous other donors. 

This work is based in part on data obtained with the {\it Einstein Probe}, a space mission supported by the Strategic Priority Program on Space Science of the Chinese Academy of Sciences, in collaboration with ESA, MPE and CNES (grant XDA15310000).
This work has made use of data from the European Space Agency (ESA) mission {\it Gaia} (\url{https://www.cosmos.esa.int/gaia}), processed by the Gaia Data Processing and Analysis Consortium (DPAC;
\url{https://www.cosmos.esa.int/web/gaia/dpac/consortium}). Funding for the DPAC has been provided by national institutions, in particular the institutions participating in the Gaia Multilateral Agreement.
This research has made use of the SIMBAD database, operated at CDS, Strasbourg, France \citep{Wenger2000}.

This work has made use of data products from the \textit{Transiting Exoplanet Survey Satellite} (TESS) mission, including the TESS Input Catalog and Candidate Target List \citep{TIC2018}, provided by the Space Telescope Science Institute (STScI) through the Mikulski Archive for Space Telescopes (MAST). This research has also utilized data from the Pan-STARRS1 DR2 catalog \citep{PS1DR22022}, which is hosted by STScI/MAST.

Some of the data presented herein were obtained at Keck Observatory, which is a private 501(c)3 nonprofit organization operated as a scientific partnership among the California Institute of Technology, the University of California, and the National Aeronautics and Space Administration. The Observatory was made possible by the generous financial support of the W. M. Keck Foundation. 
The authors wish to recognize and acknowledge the very significant cultural role and reverence that the summit of Maunakea has always had within the Native Hawaiian community. We are most fortunate to have the opportunity to conduct observations from this mountain.

A major upgrade of the Kast spectrograph on the Shane 3\,m telescope at Lick Observatory, led by Brad Holden, was made possible through gifts from the Heising-Simons Foundation, William and Marina Kast, and the University of California Observatories.
Research at Lick Observatory is partially supported by a gift from Google.
We appreciate the expert assistance of the staff of the various observatories where data were obtained.


\end{document}